\documentclass[12pt]{article}
 \usepackage{amsfonts,amsmath,amssymb}
 
\newcommand{\atopn}[2]{\genfrac{}{}{0pt}{}{#1}{#2}}
\newcommand\mb[1]{\quad\mbox{#1}\quad}
\newcommand\wt[1]{\widetilde{#1}}
\newcommand{\II}{{\mathbb I}}

\def\GG{{\mathbb G}}
\def\cB{{\cal B}}
\def\cG{{\cal G}}
\def\cL{{\cal L}} 
 \def\cM{{\cal M}} 
 \def\cN{{\cal N}} 
 \def\cS{{\cal S}}   
\def\cT{{\cal T}}
\def\cU{{\cal U}}
\def\cY{{\cal Y}} 

\def\enne{{\cal N}}

\def\tK{\widetilde{K}}

\begin{document}
\parindent=0pt
 
 \pagestyle{empty}
 
 \begin{center}
    {\Large \textbf{
     Analytical Bethe ansatz in $\mathrm{gl}(N)$ spin chains
       }}\\[1cm]
  {\large D. Arnaudon$^{(a)}$, N.~Cramp\'e$^{(b)}$, A.~Doikou$^{(a)}$, 
	\\
  L.~Frappat$^{(a)}$,
      \underline{E.~Ragoucy}$^{(a)}$}
\end{center}

\vspace{1.2em}

$^{(a)}$ Laboratoire d'Annecy-le-Vieux de Physique Th{\'e}orique
LAPTH\footnote{UMR 5108 du CNRS, associ{\'e}e {\`a} l'Universit{\'e} de
Savoie}, BP 110, F-74941 Annecy-le-Vieux Cedex, France.\\

$^{(b)}$ University of York, Department of Mathematics,
Heslington, York YO10 5DD, U.K.

\vfill\vfill

 \centerline{\textbf{Abstract}}
 We present a global treatment of the analytical
 Bethe ansatz for $\mathrm{gl}(N)$ spin chains admitting on each
 site an arbitrary representation. The method applies for closed
 and open spin chains, and also to the case of soliton
  non-preserving boundaries.
  
\vfill

  \begin{center}
\begin{tabbing}
Talk given at \=\textit{Integrable Systems}, Prague (Czech Republic), 16--18 June
2005\\
 \>\textit{Integrable Models and Applications}, EUCLID meeting,\\
\>\hspace{2.1em} Santiago (Spain), 12--16 Sept. 2005\\
\end{tabbing}
\end{center}

\vfill

\rightline{LAPTH-Conf-1117/05}
\rightline{September 2005}
 
\newpage
\pagestyle{plain}
\setcounter{page}{1}
\section{Introduction}
 The aim of this short note is to review the ``global" treatment for 
 analytical Bethe Ansatz that has been introduced in \cite{open,SNP}.
It is a unified presentation which applies to
 \underline{any} $gl(\cN)$ spin chain (whatever the quantum spaces are), and
 to  general integrable boundary condition (periodic, soliton
preserving or soliton non-preserving), provided $K^+(\lambda)=\II$. This will be achieved by the
use of only the algebraic structure described by the monodromy
matrix. 

To give a first insight to the technique we develop, we first review in
section \ref{sect-Heisenberg}
the usual closed spin chain associated to the XXX Heisenberg model
(generalized to $gl(\cN)$),
adopting a view point which will make clear the generalization to a
general closed spin chain (section \ref{sect-closed}). Then, we will
study the case of a general open spin chain, with "usual" (soliton
preserving) integrable boundary 
conditions (section \ref{sect-open}). The case of a general open spin 
chain with soliton non-preserving boundary conditions is presented in 
section \ref{sect-SNP}. Finally, we conclude in section
\ref{conclu} on open questions.

\section{Closed spin chain revisited\label{sect-Heisenberg}}
As already mentioned, the (generalized) Heisenberg spin chain has been well
studied, both from analytical and algebraic Bethe ansatz (see references in
e.g. \cite{open}). 
We just
review some of these well-known results to introduce the algebraic
set-up which will be used in the following.

The goal is to compute the spectrum of the spin chain described by the
generalized XXX $gl(\cN)$ spin chain Hamiltonian 
$H\propto \sum_{n=1}^\ell P_{n-1,n}$, which describes $\ell$ spins
interacting with their nearest neighbours.

For $gl(2)$, we recover the usual XXX Hamiltonian:
 $$H=\frac{1}{2}\sum_{n=1}^\ell 
 ~\left(~\sigma^x_{n-1}\sigma^x_{n}
+\sigma^y_{n-1}\sigma^y_{n}
 +\sigma^z_{n-1}\sigma^z_{n}
 +1~\right)\,,$$
 where $\sigma$ are the Pauli matrices, and $n$ labels the sites (with
 the site 0 identified with the site $\ell$).

The starting point is the  monodromy matrix
$T_{a}(\lambda)=R_{a1}(\lambda)R_{a2}(\lambda)\cdots
R_{a\ell}(\lambda)$,
where $a$ denotes the auxiliary space and 1,..,$\ell$ the
quantum spaces carrying the spin
variables (fundamental representation of $gl(\cN)$). 
$R_{a1}(\lambda)$ is the R-matrix of the Yangian of $gl(\cN)$,
$ R_{12}(\mu)=\mu-iP_{12}$ where $P_{12}$ is the permutation and $\mu$
the so-called spectral parameter.
In the spin chain context, it is viewed as the scattering matrix of a 
`test-particle' with a spin of the chain (see below).

The monodromy matrix obeys the  {Yangian exchange relations}:
$$R_{ab}(\lambda-\mu)\,T_{a}(\lambda)\,T_{b}(\mu)=
T_{b}(\mu)\,T_{a}(\lambda) \,
R_{ab}(\lambda-\mu)$$

These exchange relations are enough to show that the  {transfer matrix}
$t(\lambda)=tr_{a} T_{a}(\lambda)$
satisfies
$[t(\lambda),t(\mu)]=0$, that is to say
it generates (upon expansion in $\lambda$) $\ell$ commuting
(independent) quantities.
Since $H=\frac{d}{d\lambda}\ln t(\lambda)\vert_{\lambda=0}$, this proves
the {integrability} of the spin chain model.

\subsection{Analytical Bethe Ansatz \label{BAE:fundl}}
Once one has proven the integrability of the Heisenberg spin chain, a 
natural question is to look for the
spectrum of $t(\lambda)$. For this purpose, one introduces the
reference state (so-called pseudo-vacuum)
$\omega=v_{1}\otimes v_{1}\otimes\cdots\otimes v_{1}$
with $v_{1}^t=(1,0,\ldots,0)$, which
 is an eigenvector of the transfer matrix:
$$t(\lambda)\omega=\Lambda^{0}(\lambda)\omega
\mb{with} 
\Lambda^{0}(\lambda)=(\lambda+i)^\ell+(\cN-1)\,\lambda^\ell$$
Then, the major hypothesis of the analytical
Bethe ansatz states that
\underline{all} eigenvalues of the transfer matrix can be described by the
formula
$$\displaystyle \Lambda(\lambda)=(\lambda+i)^\ell
A_{0}(\lambda)+\sum_{k=1}^{\cN-1} \lambda^\ell\,A_{k}(\lambda),$$
where $A_{k}(\lambda)$ are the dressing functions, to be determined.

The determination of the dressing functions relies on the following
assumption:
the dressing functions are
``simple" rational functions with  simple poles only. Help in fixing
these dressing functions is given by the requirement that all
eigenvalues $\Lambda(\lambda)$ are analytical, i.e. the residues at each
possible "pole" vanish, and the 
fusion procedure (which provides the transfer matrix for a higher
dimensional auxiliary space starting from a $\cN$-dimensional
auxiliary space). These constraints and requirement are 
enough to deduce the form of the dressing functions:
$$A_{k}(\lambda)=\prod_{n=1}^{M_{k-1}}
\frac{\lambda-\lambda_n^{(k-1)}+\frac{i\;(k+1)}{2}}
{\lambda-\lambda_n^{(k-1)}+\frac{i\;(k-1)}{2}}
~~\prod_{n=1}^{M_{k}}
\frac{\lambda-\lambda_n^{(k)}+\frac{i\;(k-2)}{2}}
{\lambda-\lambda_n^{(k)}+\frac{i\;k}{2}}$$
where $M_{j}$ are integers and $\lambda_{n}^{(k)}$ are solution to the 
\underline{Bethe ansatz equations} (BAE)
\begin{eqnarray*}
&&\hspace{-2.1em}
\prod_{m=1}^{M_{k-1}}
{e}_{-1}\Big(\lambda_n^{(k)}\mbox{-}\lambda_m^{(k-1)}\Big)
\prod_{\atopn{m=1}{m\neq n}}^{M_{k}}
{e}_{2}\Big(\lambda_n^{(k)}\mbox{-}\lambda_m^{(k)}\Big)
\prod_{m=1}^{M_{k+1}}
{e}_{-1}\Big(\lambda_n^{(k)}\mbox{-}\lambda_m^{(k+1)}\Big)
=  \begin{cases}
    \displaystyle
    \left(e_1\left(\lambda_n^{(k)}\right)\right)^\ell \\[1.2ex]
    1
  \end{cases}
\end{eqnarray*}
where the first line corresponds to $k=1$ and the second to $k\neq1$.
We  introduced 
$\displaystyle
e_{n}(\lambda)=\frac{\lambda+\frac{in}{2}}{\lambda-\frac{in}{2}}$.
Each solution to these BAE gives a set of dressing functions, which
will themselves allow us to describe the complete spectrum of the transfer
matrix.

This method can be (and has been) applied, case by case,
for several spin chain models, so that one is led to address the
following question: can one 
extend the method to \underline{all} integrable spin chains once for
all? This is the question we answer in the following sections.

\section{``Global" treatment\label{sect-closed}}

 The key point in a treatment valid for all spin chains is the
 following remark:
the integrability and the spectrum  of the spin chain model 
rely only on the algebraic structure
of the Yangian ($RTT=TTR$ exchange relation). The way to
 tackle the problem will be in the
``translation" of all the above procedure into algebraic
properties. An immediate advantage of the procedure will be a
treatment valid for  $gl(\cN)$ spin chains whatever the quantum
spaces are. The starting point is the assumption that 
the monodromy matrix of a spin chain is a finite
dimensional irreducible
representation of the Yangian. This simple assumption will be enough
to deduce most of the properties needed for the analytical Bethe
ansatz. Indeed, the
classification of such representations tells us that one has to consider 
a product of evaluation
representations, that we first describe. 

An evaluation representation (of the Yangian) is constructed from 
irreducible representation $\pi(\alpha)$ of
$gl(\cN)$, characterized by the weight
$\alpha=(\alpha_{1},\ldots,\alpha_{\cN})$.
It is convenient to gather the represented $gl(\cN)$ generators into a single 
$\enne\times\enne$ matrix 
$$\GG=\sum_{i,j=1}^\cN\pi(\alpha)(e_{ij})\,E_{ij}\,,$$ where $E_{ij}$ are elementary
$\cN\times\cN$ matrices (with 1 in position $i,j$), $e_{ij}$ are the
$gl(\enne)$ generators, and $\pi(\alpha)(e_{ij})$ their
representation.
Then, the
representation of the Yangian generators is given by
$$
\pi_{\lambda}(\alpha)\,\cL(\lambda)=\lambda\,\II+\GG
$$
Here, it is interpreted as the scattering matrix of a test-particle (of
momentum $\lambda$ and carrying a spin in the $gl(\enne)$ fundamental 
representation) with a spin (on a site of the chain) in the
representation $\pi(\alpha)$ of $gl(\enne)$.

The monodromy matrix takes the form
$$
T_{a}(\lambda)=\pi_{\lambda+a_{1}}(\alpha^{(1)})\otimes
\pi_{\lambda+a_{2}}(\alpha^{(2)})\otimes\cdots\otimes
\pi_{\lambda+a_{\ell }}(\alpha^{(\ell )})\,
\Delta^{(\ell )}\cL(\lambda)
$$
where $a_{j}$ are free parameters (inhomogeneity parameters in spin
chain context). It describes the response of the whole spin chain to
the test-particle.

This form leads us to introduce the notion of a 
 "global'' (unrepresented) monodromy matrix
$$\cT_{a}(\lambda)=\Delta^{(\ell )}\cL(\lambda)
=\cL_{a1}(\lambda)\cL_{a2}(\lambda)\cdots\cL_{a\ell}(\lambda)$$ 
It is easy to show that this global monodromy matrix
obeys the Yangian exchange relations as soon as $\cL(\lambda)$ does.
Correspondingly, we introduce a 
 {"global'' transfer matrix }
$$t(\lambda)=tr_{a}\cT(\lambda).$$
The Yangian exchange relations  show 
that $[t(\lambda),t(\mu)]=0$: one gets the
integrability of the model. Moreover,
the exchange relations of the monodromy matrix imply that the
Lie algebra $gl(\cN)$ commutes with $t(\lambda)$:  the model has a 
$gl(\cN)$ symmetry.

\subsection{Spectrum and general structure of the BAE}
Now that the integrability of the model is established, we turn to the
determination of the spectrum of the transfer matrix, within the
framework of analytical Bethe ansatz. For such a purpose, one has to
characterize more precisely the irreducible finite representations of 
the Yangian. They are characterized by the notion of Drinfel'd
polynomials: a  representation is associated by (i.e. in one-to-one 
correspondence with) the  {Drinfel'd polynomials}
$$P_k(\lambda)=\prod_{n=1}^\ell
\left(\lambda+a_{n}-\alpha^{(n)}_k\right)\,,\ k=1,..,\cN-1$$
The weight $(\mu_{1}(\lambda),\ldots,\mu_{\cN}(\lambda))$
of the representation  is then determined through the relations
$$
\frac{\mu_{i}(\lambda)}{\mu_{i+1}(\lambda)}=
\frac{P_{i}(\lambda)}{P_{i}(\lambda+1)}
$$
The reference state (pseudo-vacuum) is the highest vector $\omega$ 
of the representation, whose eigenvalue under the transfer matrix is 
$\displaystyle\Lambda^0(\lambda)=\sum_{k=1}^\cN ~P_k(\lambda).$

Then one assumes the following form for all the eigenvalues:
$$\displaystyle \Lambda(\Lambda)=\sum_{k=1}^\cN ~
P_k(\lambda)\;A_k(\lambda).$$
The dressing functions 
 $A_{k}(\lambda)$ are supposed to be ``simple" rational functions 
 with simple poles.
Requiring analyticity of the eigenvalues, and using fusion (which in
the present algebraic context amounts to the existence of a center in 
the Yangian) fixes the form of the dressing functions
$$
A_{k}(\lambda)=\prod_{n=1}^{M_{k-1}}
\frac{\lambda-\lambda_n^{(k-1)}+\frac{i\;(k+1)}{2}}{\lambda-\lambda_n^{(k-1)}-\frac{k-1}{2}}
\prod_{n=1}^{M_k}
\frac{\lambda-\lambda_n^{(k)}+\frac{i\;(k-2)}{2}}{\lambda-\lambda_n^{(k)}-\frac{k}{2}}
$$
where the parameters $\lambda_{n}^{(k)}$ obey the Bethe Ansatz
Equations (BAE):
 \begin{eqnarray*}
&&
 \hspace{-2.1em}
\prod_{m=1}^{M_{k-1}}
{e}_{-1}\Big(\lambda_n^{(k)}\mbox{-}\lambda_m^{(k-1)}\Big)
\prod_{\atopn{m=1}{m\neq n}}^{M_{k}}
{e}_{2}\Big(\lambda_n^{(k)}\mbox{-}\lambda_m^{(k)}\Big)
\prod_{m=1}^{M_{k+1}}
{e}_{-1}\Big(\lambda_n^{(k)}\mbox{-}\lambda_m^{(k+1)}\Big)
=   \frac{P_{k}\left(\lambda_n^{(k)}-\frac{i\;k}{2}\right)}
{P_{k+1}\left(\lambda_n^{(k)}-\frac{i\;k}{2}\right)} \qquad\quad
\end{eqnarray*}
This reproduces the results found by other methods \cite{KuRe,OgWi}.
Let us remark that the left-hand side of the BAE reflects the
Lie algebra dependence (through the
Cartan matrix of $gl(\cN)$), while their
right-hand side shows up a
representation dependence (through Drinfel'd
polynomials).

Note that the choice of a closed spin chain model
is fixed by the choice of the quantum spaces, i.e. the choice
of the Drinfel'd polynomials $P_{k}(\lambda)$. Once these polynomials 
are given, the spectrum of the transfer matrix is fixed through the
resolution of the BAE.

We give hereafter some examples (more can be
found in \cite{open}).

\subsection{Examples}
\subsubsection{Closed spin chain in the fundamental
  representation\label{sect:usual}} 
The usual closed spin chain corresponds to spins in the fundamental
representation. The Hamiltonian is given by the well-known formula
\begin{eqnarray}
  \label{Hln}
  H=\left.\frac{d}{d\lambda}\;\big( ln~
  t(\lambda)\big)\right|_{\lambda=0}\,.
\end{eqnarray}
In this case, we have $\alpha^{(n)}=(1,0,\dots,0)$, for $1\leq n \leq
\ell$. Then, the Drinfel'd polynomials read
\begin{eqnarray}
  P_k(\lambda)=
  \begin{cases}
    \displaystyle
    \prod_{j=1}^\ell(\lambda+a_j+i)&,~~ k=1\\
    \displaystyle
    \prod_{j=1}^\ell(\lambda+a_j)&,~~ k\neq 1
  \end{cases}
\end{eqnarray}
Plugging these expressions in the Bethe equations, we 
recover the usual Bethe equations for closed spin
chains, as given in section \ref{BAE:fundl}.

\subsubsection{Closed spin chain for non-fundamental representations
\label{sect:sspin}}

One can generalize the above example to the case where all the spins 
belong to the same (not necessarily fundamental) representation, 
given by 
\begin{equation}
    \alpha^{(1)}=\alpha^{(2)}=\ldots=\alpha^{(\ell)}=
    (\alpha_{1},\alpha_{2},\dots,\alpha_{\enne})\,.
\end{equation}
This leads to the following Drinfel'd polynomials
\begin{eqnarray}
  P_k(\lambda)=(\lambda+i\alpha_{k})^\ell \ \mbox{ so that }\ 
  \frac{P_{k}\left(\lambda_n^{(k)}-\frac{i\;k}{2}\right)}
  {P_{k+1}\left(\lambda_n^{(k)}-\frac{ir\;k}{2}\right)} =
  \left[e_{\beta^-_{k}}
  \left(\lambda_n^{(k)}-i\frac{k-\beta_{k}^{+}}{2}\right)\right]^\ell
  \;,
\end{eqnarray}
with $\beta_{k}^\pm=\alpha_{k}\pm\alpha_{k+1}$.
In particular, we recover the result given in
\cite{KRSTBFK} about the XXX higher spin chains.

\section{Open spin chains and reflection algebra\label{sect-open}}
The technique can be applied to the case of open spin chains. Here, the
algebraic structure, instead of being the 
Yangian algebra $Y(\cN)$, is the  \underline{reflection
algebra} \cite{Cher}
$\cB(\cN,\cM)$, whose generators (gathered into a matrix
$\cB(\lambda)$) obey the exchange relations (reflection equation):
\begin{eqnarray*}
&&\hspace{-2.1em}  R_{ab}(\lambda_{a}-\lambda_{b})\,
\cB_{a}(\lambda_{a})\,
  R_{ba}(\lambda_{a}+\lambda_{b})\, \cB_{b}(\lambda_{b})	=
  \cB_{b}(\lambda_{b})\,
  R_{ab}(\lambda_{a}+\lambda_{b})\, \cB_{a}(\lambda_{a})\,
  R_{ba}(\lambda_{a}-\lambda_{b}) 
\end{eqnarray*}
The reflection algebra is a coideal subalgebra of the Yangian, so that
the ``global" open spin chain monodromy matrix takes the form:
\begin{eqnarray*}
\cB_a(\lambda) &=&
\cL_{a1}(\lambda)\;\dots\;\cL_{a\ell}(\lambda)~
K_a(\lambda)~
\cL_{a\ell}^{-1}(-\lambda)\;\dots\;\cL_{a1}^{-1}(-\lambda)\\
& =&\Delta^{(\ell)}B(\lambda)
=\cT_{a}(\lambda)\,K_{a}(\lambda)\,\cT_{a}(-\lambda)^{-1}
\end{eqnarray*}
where $K(\lambda)$ is a matricial solution to the reflection equation.

The
classification of such matrices has been done \cite{Mint}, it
implies that diagonalisable matrices 
\underline{must}
be of the form ($\xi$ is a free parameter):
\begin{eqnarray*}
K(\lambda) = U\,\left(\,{\lambda}\,\mathbb
 E_{\cM}\,+\,{\xi}\,\II_\cN\,\right)U^{-1}
 \equiv U\, D_{\xi}(\lambda)\,U^{-1}\,,\\
 \mbox{ with }\
\mathbb E_{\cM} = diag(\underbrace{1,\ldots,1}_{\cM},
\underbrace{-1,\ldots,-1}_{\cN-\cM})\,.
\end{eqnarray*}
The
complete classification of $\cB(\cN,\cM)$
finite dimensional irreducible
representations  has been done in \cite{momo} and
uses the evaluation representations presented in previous
section.

As in the closed spin chain case, one defines a
"global'' transfer matrix
$$b(\lambda)=tr_a\left(\cB_a(\lambda)\right)$$
which contains the Hamiltonian
$$ H = -\frac{1}{2} \frac{d}{d \lambda}b(\lambda)\Big\vert_{\lambda
=0}.$$

Again, the underlying algebraic structure is sufficient to prove
$$[b(\lambda),b(\mu)]=0,$$
i.e. the  {integrability} of the
models, and the $gl(\cM)\oplus gl(\cN-\cM)$ Lie
algebra symmetry.

Its spectrum depends on $K(\lambda)$ \underline{only} through
$D_{\xi}(\lambda)$.

Note that a more general treatment
can be done using a $K^{+}(\lambda)$ matrix solution to a dual reflection
equation, the transfer matrix being
$b(\lambda)=tr_{a}K_{a}^{+}(\lambda)\cB_{a}(\lambda)$. The present
techniques still applies if one supposes that $K(\lambda)$ and $K^{+}(\lambda)$
commute (for instance when they are both diagonal), see \cite{open}.
The case presented here corresponds to $K^+(\lambda)=\II$ (which is
indeed a solution to the dual reflection equation).

\subsection{BAE for open spin chains}
The pseudo-vacuum is the
highest weight vector of the representation, with
eigenvalue 
$$\displaystyle\Lambda^0(\lambda)=
\sum_{k=1}^\cN g_k(\lambda)~\beta_k(\lambda),$$
where $\beta_{k}(\lambda)$ play the role of Drinfel'd polynomials for
reflection algebra and
$$g_k(\lambda)=
\frac{2\lambda(2\lambda+i\cN)}
{(2\lambda+ik-i)(2\lambda+ik)}D_{kk}(\lambda)
$$
correspond to the chosen matrix $K(\lambda)=UD(\lambda)U^{-1}$.

The form of the eigenvalues is assumed to be
$$\displaystyle \Lambda(\lambda)=\sum_{k=1}^\cN
g_k(\lambda)~\beta_k(\lambda)~A_k(\lambda)
,$$
with dressing functions $A_{k}(\lambda)$ to be determined.

Analyticity and fusion lead to
\begin{eqnarray*}
A_{k}(\lambda)&=&
\prod_{n=1}^{M_{k-1}}
\frac{\lambda+\lambda_n^{(k-1)}+\frac{i(k+1)}{2}}
{\lambda+\lambda_n^{(k-1)}+\frac{i\;(k-1)}{2}}
\quad
\frac{\lambda-\lambda_n^{(k-1)}+\frac{i(k+1)}{2}}
{\lambda-\lambda_n^{(k-1)}+\frac{i\;(k-1)}{2}}
\\
&&\times 
\prod_{n=1}^{M_k}
\frac{\lambda+\lambda_n^{(k)}+\frac{ik}{2}-i}
{\lambda+\lambda_n^{(k)}+\frac{i\;k}{2}}
\quad
\frac{\lambda-\lambda_n^{(k)}+\frac{ik}{2}-i}
{\lambda-\lambda_n^{(k)}+\frac{i\;k}{2}}
\end{eqnarray*}
where the $\lambda_{n}^{(k)}$ parameters obey the  {BAE:
\begin{eqnarray*}
&&\prod_{m=1}^{M_{k-1}}
\wt{e}_{-1}\left(\lambda_n^{(k)},\lambda_m^{(k-1)}\right)
\prod_{\atopn{m=1}{m\neq n}}^{M_{k}}
\wt{e}_{2}\left(\lambda_n^{(k)},\lambda_m^{(k)}\right)
\prod_{m=1}^{M_{k+1}}
\wt{e}_{-1}\left(\lambda_n^{(k)},\lambda_m^{(k+1)}\right)
\\
&&\qquad\qquad=
\frac{\beta_{k}\left(\lambda_n^{(k)}-\frac{i\;k}{2}\right)}
{\beta_{k+1}\left(\lambda_n^{(k)}-\frac{i\;k}{2}\right)}
\times
\begin{cases}
\ -e_{-\cM-2i\xi}\left(\lambda_m^{(\cM)}\right)
\mb{if $k=\cM$} \\
\ 1 \qquad\qquad\qquad\qquad\mb{otherwise} 
\end{cases}
\end{eqnarray*}}
where 
$
\wt{e}_{n}(\lambda,\mu)=e_{n}\left(\lambda-\mu\right)
e_{n}\left(\lambda+\mu\right)$.

Again, the left-hand side is linked to the $gl(\cN)$ Cartan matrix,  while
the right-hand side is related to the chosen representations.

The choice of an open spin chain model is now determined by two
types of data:

1- The
choice of the quantum spaces, i.e. the choice
of the Drinfel'd polynomials $\beta_{k}(\lambda)$\\
2- The choice of the boundary condition, i.e. the choice of $K(\lambda)$
(in fact the eigenvalues of $K(\lambda)$), which fixes
$g_{k}(\lambda)$.

Then, the spectrum of the transfer matrix is given by the solutions to
the BAE.

\section{Soliton non-preserving open spin chains\label{sect-SNP}}
We now turn to the case of integrable spin chains with a boundary such 
that a soliton reflects into
an anti-soliton (and vice-versa) \cite{doikou1}.
The underlying algebraic structure is now the  \underline{twisted Yangian}
\cite{Olsh,MNO}, whose
exchange relations are ($\rho=-\frac{\cN}{2}$):
\begin{eqnarray*}
&&\hspace{-2.1em}
R_{ab}(\lambda_{a}-\lambda_{b})\, {\cS}_{a}(\lambda_{a})\,
  R_{ab}^{t_a}(-\lambda_{a}-\lambda_{b}+i\rho)\,
{\cS}_{b}(\lambda_{b})\\
&& =
  {\cS}_{b}(\lambda_{b})\,
  R_{ab}^{t_a}(-\lambda_{a}-\lambda_{b}+i\rho)\,
{\cS}_{a}(\lambda_{a})\,
  R_{ba}(\lambda_{a}-\lambda_{b})\quad
\end{eqnarray*}

The  {``global" monodromy matrix} reads
$ \cS_a(\lambda)=\cT_a(\lambda)\,{\tK}_a
\,{\cT_a}^{t_a}(-\lambda+i\rho)=\Delta^{(\ell)}S(\lambda)$, where
$\tK_{a}(\lambda)$ is a matricial solution to the twisted Yangian
exchange relations. It can be proven that it must be a constant matrix
$\tK_{a}$.

The ``global" transfer matrix is now\footnote{A more general treatment
can be done using a $\tK^{+}(\lambda)$ matrix solution to a dual exchange
relation, the transfer matrix being
$s(\lambda)=tr_{a}\tK_{a}^{+}(\lambda)\cS_{a}(\lambda)$, see \cite{SNP}.}
$s(\lambda)=tr_{a}\cS_{a}(\lambda)$.

Performing the same 
 construction starting from the highest weight vector and using the
 classification of the twisted Yangian finite dimensional irreducible representations
 \cite{twmolev}, we get the
 following form for the eigenvalues
$$\displaystyle
\Lambda^0(\lambda)= \sum_{k=1}^\cN
    g_k(\lambda)~\sigma_k(\lambda) \qquad;\qquad
\Lambda(\lambda)= \sum_{k=1}^\cN
    g_k(\lambda)~\sigma_k(\lambda)~A_k(\lambda)
,$$
where $\sigma_{k}(\lambda)$ are the Drinfel'd polynomials for the
twisted Yangian and $g_{k}(\lambda)$ are related to the matricial solution
$\tK_{a}(\lambda)$.

Analyticity and fusion allow us to determine the
dressing functions and to get the BAE. The generic form of these
laters being (see \cite{SNP} for the complete set):
\begin{eqnarray*}
&&\hspace{-2.1em}
\prod_{m=1}^{M_{k-1}}
\wt{e}_{-1}\left(\lambda_n^{(k)},\lambda_m^{(k-1)}\right)
\prod_{\atopn{m=1}{m\neq n}}^{M_{k}}
\wt{e}_{2}\left(\lambda_n^{(k)},\lambda_m^{(k)}\right)
\prod_{m=1}^{M_{k+1}}
\wt{e}_{-1}\left(\lambda_n^{(k)},\lambda_m^{(k+1)}\right)\\
&&=\frac{g_k(\lambda_p^{(k)}+\frac{\hbar k}{2})
\sigma_k(\lambda_p^{(k)}+\frac{\hbar k}{2})}
{g_{k+1}(\lambda_p^{(k)}+\frac{\hbar k}{2})
\sigma_{k+1}(\lambda_p^{(k)}+\frac{\hbar k}{2})} 
\end{eqnarray*}

\section{Conclusion and perspectives\label{conclu}}

We have formulated a ``global" treatment for analytical Bethe Ansatz for
 {any} $gl(\cN)$ spin chain (whatever the quantum spaces are), and
applicable to general integrable boundary condition (periodic, soliton
preserving or soliton non-preserving), provided $K^+(\lambda)=\II$.

This proves the integrability
of the spin chains on general ground, allows us to compute their
symmetry algebra and leads to the exact form of
BAE for all these spin chains.

It remains to find, within this formalism, an expression for a
local Hamiltonians. Indeed,  although the BAE provide the spectrum of 
the transfer matrix (hence of all Hamiltonians), local Hamiltonians are obtained using 
higher dimensional auxiliary spaces, so that their expression in term 
of global transfer matrix is still lacking.
 
An interesting point is also the question of the completeness of the
spectrum, which seems to be  equivalent to the irreducibility of the
representation. It is at least a necessary condition.
 
Finally, generalization of this approach to other algebras should be
done: the case of  $U_{q}(\widehat{gl}(\cN))$  is under
investigation, while the case of super-Yangian $\cY(gl(\cN\vert\cM))$ 
seems to be an easy task, although indecomposable
representations may lead to new problems\footnote{We thank V. Rittenberg
for drawing our attention to this point.}.

\section*{Acknowledgments}
This work has been financially supported by the TMR Network
EUCLID: ``Integrable models and applications: from strings to
condensed matter'', contract number HPRN-CT-2002-00325.


\begin{thebibliography}{9}
\bibitem{open}  {D.~Arnaudon, N.~Cramp\'e, A.~Doikou, L.~Frappat
and {\'E}.~Ragoucy}, 
 \textsl{Analytical Bethe Ansatz
for closed and open $gl(N)$-spin chains in any representation},
J. Stat. Mech. (2005) P02007 and \texttt{math-ph/0411021}.

\bibitem{SNP} {D.~Arnaudon, N.~Cramp\'e, A.~Doikou, L.~Frappat
and {\'E}.~Ragoucy}, 
\textsl{Analytical Bethe Ansatz for open spin chains with soliton 
non-preserving boundary conditions}, 
\texttt{math-ph/0503014}.
	 
\bibitem{KuRe} P. Kulish and N. Yu Reshetikhin,   
\textsl{Diagonalisation of $GL(N)$ invariant transfer matrices and quantum 
N-wave system (Lee model)}, 
J. Phys. \textbf{A16} (1983) L591.

\bibitem{OgWi}
E. Ogievetsky and P. Wiegmann,
\textsl{Factorized S-matrix and the Bethe ansatz for simple Lie
 groups,} 
 Phys. Lett. \textbf{B168} (1986) 360.

\bibitem{Cher} I.V.~Cherednik, 
\textsl{Factorizing particles on a half line and root
 systems,} 
Theor. Math. Phys. \textbf{61} (1984) 977.

\bibitem{Mint} M.~Mintchev, E.~Ragoucy and P.~Sorba,
\textsl{Spontaneous symmetry breaking in the gl(N)-NLS hierarchy on 
the half line},
J.\ Phys. {\bf A34} (2001) 8345 and \texttt{hep-th/0104079}.

\bibitem{momo} A. I. Molev and E. Ragoucy,
\textsl{Representations of reflection algebras},
Rev. Math. Phys. {\bf 14} (2002) 317 and \texttt{math.QA/0107213}.
  
\bibitem{doikou1} A.~Doikou,
\textsl{Quantum spin chain with ``soliton non preserving''
 boundary conditions,} 
J. Phys. \textbf{A33} (2000) 8797 and \texttt{hep-th/0006197}.

\bibitem{Olsh} G.I.~Olshanski,
 \textsl{Extension of the algebra $\cU(\cG)$ for infinite dimensional 
 classical Lie algebras $\cG$, and the Yangians $\cU(gl(m))$},
Soviet. Math. Dokl. \textbf{36} (1988) 569; \\
\textit{id.},
 \textsl{Representations of infinite dimensional classical groups, 
 limits enveloping algebras and Yangians}, 
 in``Topics in Representation theory", A.A.~Kirillov ed., 
Adv. in Soviet. Math. \textbf{2} (1998) 1.
  
\bibitem{MNO}
A.~Molev, M.~Nazarov and G.~Olshanski,
\textsl{Yangians and classical Lie algebras}, 
Russian Math. Survey \textbf{51} (1996) 205 and
  \texttt{hep-th/9409025}.

\bibitem{twmolev} A. I. Molev,
 \textsl{Finite-dimensional irreducible representations of twisted
 Yangians,} 
J. Math. Phys. \textbf{39} (1998) 5559 and \texttt{q-alg/9711022}.

\bibitem{KRSTBFK} P.P. Kulish, N.Yu. Reshetikhin and E.K. Sklyanin,
  \textsl{Yang-Baxter equation and representation theory: I},
  Lett. Math. Phys. \textbf{5} (1981) 393;\\
 L.A. Takhtajan, 
  \textsl{The picture of low-lying excitations in the isotropic
    Heisenberg chain of arbitrary spins},
  Phys. Lett. \textbf{A87} (1982) 479;\\
 H.M. Babujian, \textsl{Exact solution of the isotropic 
    Heisenberg chain with arbitrary spins: thermodynamics of the model}, 
  Nucl. Phys. \textbf{B215} (1983) 317;\\
 L.D. Faddeev, \textsl{How Algebraic Bethe Ansatz works
    for integrable model}, Les Houches summerschool 1995 and
  \texttt{hep-th/9605187};\\
  N.~Kitanine, \textsl{Correlation functions of the
    higher spin XXX chains},
  \texttt{math-ph/0104016}.
\end{thebibliography}
\end{document}